\documentclass[%
 aps,
 PRB,%
 amsmath,amssymb,
 reprint,%
]{revtex4-1}

\usepackage{graphicx}
\usepackage{bm}
\usepackage{caption}
\usepackage{subcaption}
\usepackage{afterpage}
\usepackage{epstopdf}
\usepackage{amsmath}
\usepackage{etoolbox}

\def \BT {Bi$_{2}$Te$_{3}$}
\def \BS {Bi$_{2}$Se$_{3}$}
\def \BTS {Bi$_{2}$Te$_{2}$Se}
\def \wavnum {cm$^{-1}$}
\def \Bicompound {Bi$_{2}$Te$_{3-x}$Se$_{x}$}
\newcommand{\DM}{V$_{1}$ mode}

\newcommand{\all}{$et$ $al$.}
\newcommand{\BSCO}{Bi$_{2}$Sr$_{2}$CuO$_{6}$}

\begin{document}

\preprint{AIP/123-QED}

\title{Local Phonon Mode in thermoelectric \BTS{} from charge neutral antisites}

\author{Yao Tian}
\affiliation{Department of Physics \& Institute of Optical Sciences, University of Toronto, ON M5S 1A7, Canada}

\author{Gavin B. Osterhoudt}
\affiliation{Department of Physics, Boston College 140 Commonwealth Ave Chestnut Hill, MA 02467-3804, USA}

\author{Shuang Jia}
\thanks{Current address: School of Physics, Peking University, Beijing China 100871}
 \affiliation{Department of Chemistry, Princeton University, Princeton, NJ 08540, USA}




\author{R. J. Cava}
\affiliation{Department of Chemistry, Princeton University, Princeton, NJ 08540, USA}

\author{Kenneth S. Burch}
\email{ks.burch@bc.edu}
\affiliation{Department of Physics, Boston College 140 Commonwealth Ave Chestnut Hill, MA 02467-3804, USA}

\date{\today}

\begin{abstract}
Local modes caused by defects play a significant role in the thermal transport properties of thermoelectrics. Of particular interest are charge-neutral defects that suppress thermal conductivity, without significantly reducing electrical transport. Here we report a temperature dependent Raman study, that identifies such a mode in a standard thermoelectric material, \BTS{}. One of the modes observed, whose origin has been debated for decades,  was shown mostly likely  to be an antisite defect induced local mode. The anomalous temperature independent broadening of the local mode is ascribed to the random arrangement of Se atoms. The temperature renormalization of all modes are well explained by an anharmonic model--Klemens's model.
\end{abstract}

\maketitle

%

The family of compounds, \Bicompound{}  have been studied for decades as   good thermoelectrics.\cite{goldsmid2009introduction} In binary compounds like \BS{} and \BT{}, the small defect formation energies and  band gaps often result in relatively large carrier densities, leading the bulk conductance to dominate over
the surface conductance.\cite{jia2011low} A transition from p-type to n-type behavior in the
\Bicompound{} solid solution was reported decades ago.\cite{sokolov2004chemical,Nakajima1963479}
Further studies have revealed that when x is close to 1, namely \BTS{}, the crystal structure is ordered.\cite{jia2011low}  The ordering was suggested to produce the dramatic reduction in defect density.\cite{PhysRevB.82.241306} 
However, surprisingly it has been known that \BTS{} contains  a single extra mode, the origin of which remains a mystery and was suspected to be a local mode.\cite{richter1977raman}  It is known that local modes play an important role in lowering the thermal conductivity of materials. Especially for thermoelectric clathrates\cite{takabatake2014phonon} and skutterudites,\cite{PhysRevB.89.184304}  intentionally induced local modes (rattling modes) can cause the glassy behavior in phonon transport, while producing little, if any effect on the electronic properties namely phonon-glass electron-crystal.
Raman spectroscopy, owing to non-destructive to samples and quick access to phonon modes, has been widely used for characterization of materials for decades. Moreover,  combined with temperature dependence Raman scattering can be used to probe
phonon dynamics,\cite{compatible_Heterostructure_raman,Raman_Characterization_Graphene,tian2014polarized,Raman_graphene} as well as defects and defects induced local modes.\cite{ferrari2013raman,reich2004raman,barker1975optical} While there have been many studies on \Bicompound{} using Raman spectroscopy,\cite{richter1977raman,Raman_Temperature_Bi2Se3} a high spectral resolution broad temperature range dependent study is still lacking.

In this paper, we present a Raman study of \BTS{} over the temperature range from 10 K to 290 K.  Four modes were observed in the whole temperature range. Three of them are assigned as A$_{g}^{1}$, E$_{g}$ and A$_{g}^{2}$ modes.\cite{richter1977raman} The extra mode, the origin of which is controversial,  is shown most likely to be an antisite defects induced local mode. The anomalous broadness of the local mode is qualitatively understood through a simulation of diatomic chains.  The temperature dependence  of all four phonon modes are well explained by anharmonic phonon-phonon interactions.

The \BTS{} single crystal was grown with special techniques to suppress carrier-concentration. 
Detailed growth procedure and characterizations are described in previously published works.\cite{jia2011low,FTIR_Be2Te3, zareapour2014evidence}
As the preparation for measurements, \BTS{} sample  was freshly cleaved  and quickly placed inside a sample chamber. 
The temperature dependence was achieved via an automated closed-cycle cryostation manufactured by Montana Instrument, Inc. The Raman spectra were taken in a backscattering configuration with a home-built Raman microscope.  A linear polarized 532nm solid state laser was used as the excitation source.
Two diffractive Notch Filters were used to reject Rayleigh scattering. This also allows us to observe both Stokes and anti-Stokes Raman shifts.  The laser spot size was 1 $\mu{}$m in diameter. The laser power was kept as low as 40 $\mu$W to avoid laser-induced heating. This was checked at 10 K by monitoring the anti-Stokes signal as the laser power was reduced. Once the anti-Stokes signal disappeared, the power was cut an additional ≈ 50\%.  More details about the instruments can be found in elsewhere.\cite{sandilands2010stability,beekman2012raman,sandilands2015scattering,RSI_unpublished}

\begin{figure}
   \includegraphics[width=\columnwidth]{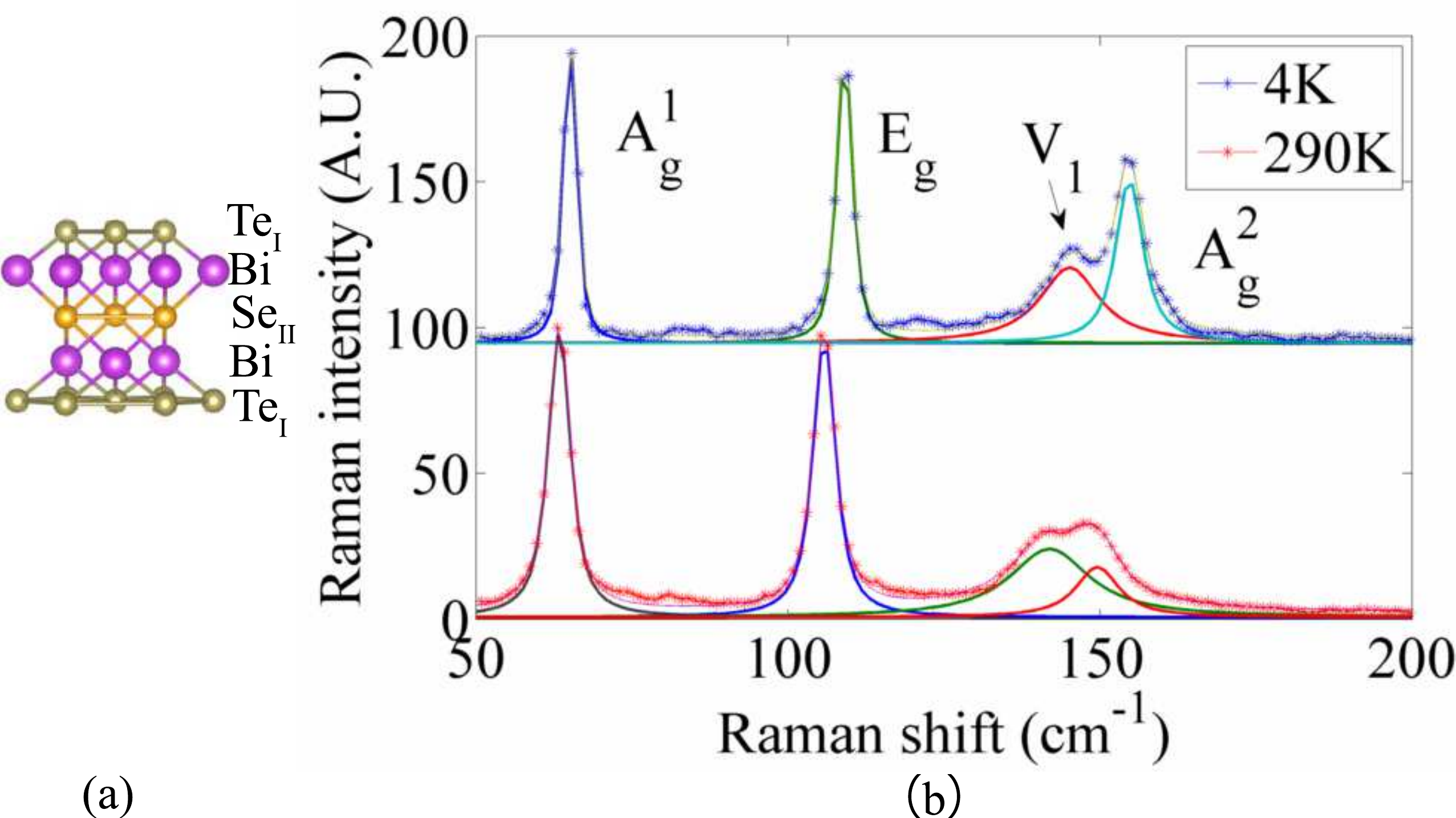}\\
  \caption{a:Schematic crystal structures of \BTS{}. b:Raman spectra of \BTS{} taken at 4 K and 290 K under XX (colinear polarzied) configuration. Raw data are shown by * markers. The four individual Voigt functions are shown in different colors. The added curves are show in thinner line.}\label{fig:BTS_raman_raw_fit}
\end{figure}

\BTS{} is of a quintuple layer structure Te$_{I}$-Bi-Se$_{II}$-Bi-Te$_{I}$ (see FIG. \ref{fig:BTS_raman_raw_fit}a).\cite{jia2011low}
From group theory analysis, the space group of \BTS{}  is R-3m (No.166). The point group is D$_{3d}$. There are four inequivalent irreducible representations in  the D$_{3d}$ point group. They are A$_{g}$, E$_{g}$,  A$_{u}$ and E$_{u}$  respectively.
Raman-active modes are 2A$_{g}$ + 2E$_{g}$. 
As established in previous experiments, the lowest energy mode in the \Bicompound{} is typically below 50 \wavnum{} which is below our low energy cut-off.\cite{PhysRevB.84.195118,hellman2014phonon,PhysRevApplied.3.014004} As a result, in our spectral range we expect to observe three phonon peaks from the zone center modes in the Raman spectra of \BTS{}.

\begin{figure}
\centering
\includegraphics[width=\columnwidth]{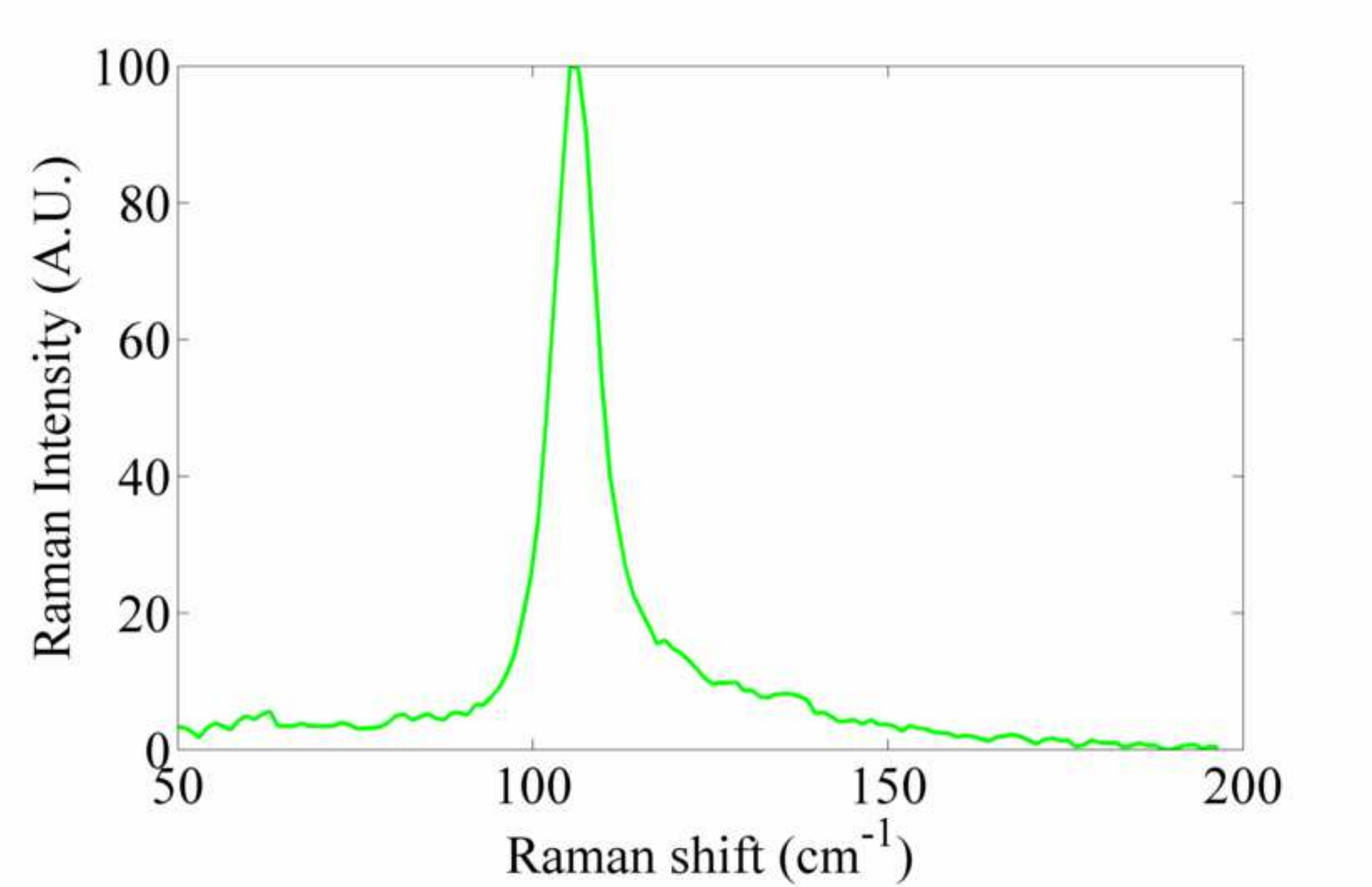}
\caption{Raman spectra of \BTS{} under XY (crossed polarized) configuration at room temperature.}
\label{fig:BTS_XX_XY}
\end{figure}

\begin{figure}
  \begin{center}
  \includegraphics[width=\columnwidth]{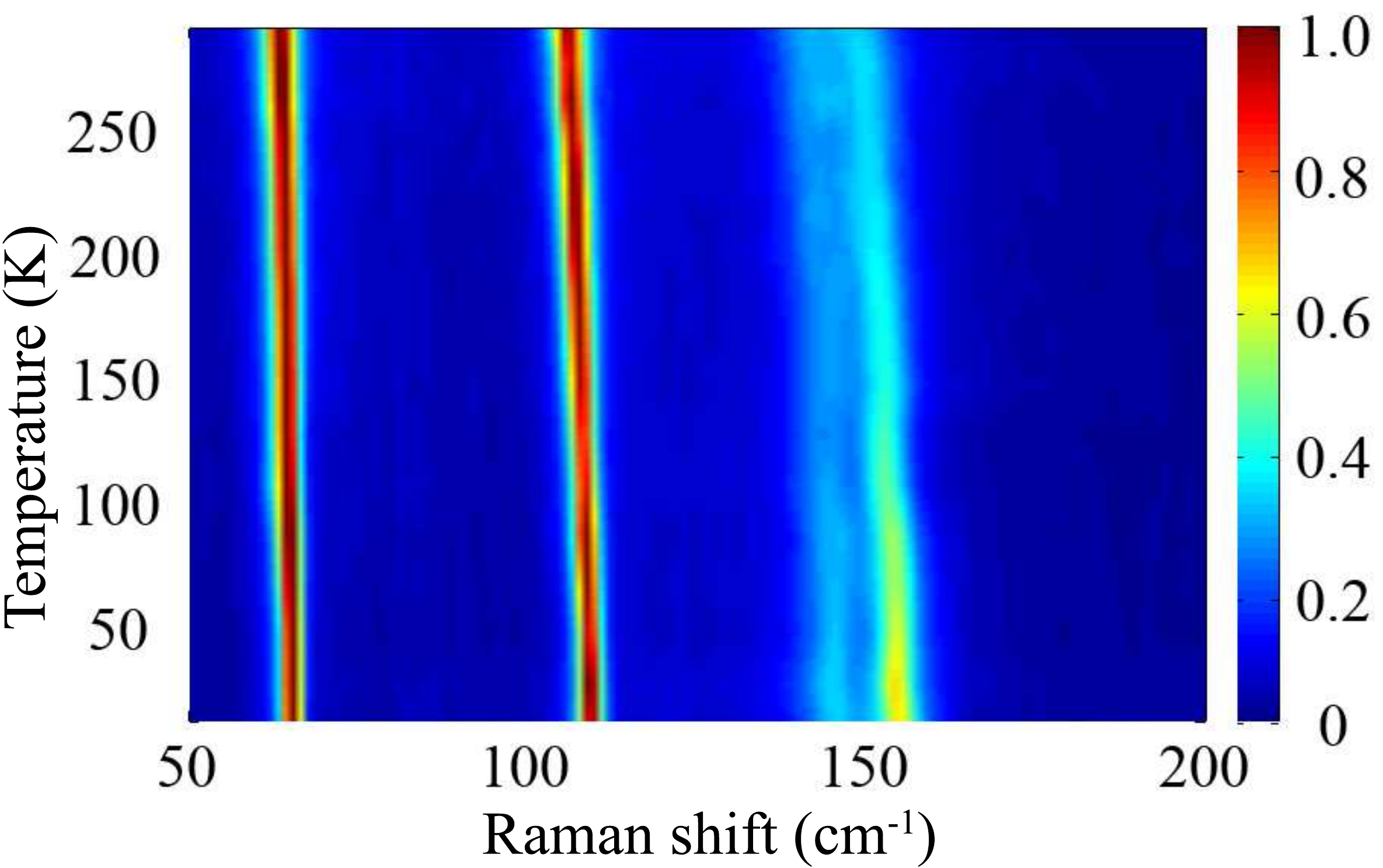}
  \end{center}
  \caption{The temperature dependent  Raman spectra of \BTS.}\label{fig:BTS_raman_colorplot}
\end{figure}

The spectra of \BTS {} taken at 290 K and 10 K under XX configuration (collinear polarized) are shown in FIG. \ref{fig:BTS_raman_raw_fit}b. We can see  the \BTS{} spectra consists of two isolated peaks at lower frequencies, and two peaks at higher energies which overlap. To  quantitatively evaluate the overlapped  features, the Raman spectra of \BTS{} were fitted with Four Voigt functions (more details about Voigt function are given later). From the fit, the  frequencies of the four features are: 65.12 \wavnum{}, 109.1\wavnum{}, 145.0 \wavnum{} and 154.6 \wavnum{} at 10 K, in good agreement with those found by Akrap \all{}.\cite{akrap2012optical} However, the line-width in our spectra are much narrower, perhaps due to the higher quality of our crystal.  In FIG. \ref{fig:BTS_XX_XY}, we also show the Raman spectra of \BTS{} under XY (crossed polarized) configuration at room temperature. It is clear that all the modes except the one located at 109.1\wavnum{} vanish in XY configuration which is consistent with the literature\cite{richter1977raman}.  The first two modes plus the mode of highest energy have been assigned as A$_{g}^{1}$ , E$_{g}$  and A$_{g}^{2}$ modes according to the literature.\cite{richter1977raman}
However, there is no consensus on the assignment of the mode at 145.0 \wavnum, which we denote as V$_{1}$ for convenience. Akrap \all{} noted that a very weak mode seen in their IR conductivity spectra was very close to the \DM{}. From this the authors concluded that the appearance of this mode in Raman scattering is probably due to the activation of A$_{2u}$ mode by disorder and symmetry breaking. 
Given the similar energy and temperature induced broadening of phonon linewidth from 10 K to 290 K ( 3 \wavnum{} broadening for \DM{}, and 3.6 \wavnum{}  for A$_{g}^{2}$ mode), one would expect a similar temperature independent broadening at low temperature.
However as described below we found that upon cooling to 10 K, the A$_{g}^{2}$ mode becomes much narrower, as is typically observed, while the \DM{} remains considerably broad. Perhaps more problematic, disorder or symmetry-breaking induced IR modes in Raman spectra are typically very weak\cite{zhao2011fabrication} and cannot be as strong as ordinary Raman modes (especially when one compares  the \DM{} to the A$_{g}^2$ mode at room temperature).

Richter \all{}\cite{richter1977raman} also observed this mode and suspected it to be a splitting of the A$_{g}^{2}$ mode resulting from the local change in bonds between the different chalcogenide layers (ie. Se in the middle of the quintuple versus Te in the outer layers). However, these measurements were performed at room temperature and thus could not tell whether this was indeed a local mode. Defects could also enable non-zone center phonons to appear in the Raman spectra, as is typically seen in graphite (i.e. the D-band).  This mechanism can generate a broad Raman profile due to the participation of multiple modes.\cite{ferreira2010evolution} This phenomenon  is usually related to electronic resonance effect.
Nonetheless, this possibility also seems unlikely, when comparing \BTS{} to \BT{}. Specifically \BT{} has of a large density of Bi$_{Te}$ antisite defects that generally lead to its heavy p-type doping.\cite{jia2012defects,jia2011low}.  Considering the high similarity in both electronic and phonon structure between  \BT{} and \BTS{} , one should also observe a ``V$_{1}$-like'' mode in \BT.\cite{PhysRevApplied.3.014004} Nonetheless despite extensive Raman measurements by various groups, the \DM{} has only been observed in samples with Se doping.\cite{richter1977raman,Bi_family_unpublished}

The comparison with \BT{} suggests we carefully consider the role of Se in generating this new Raman mode.  In theory, Se  enters the Se$_{II}$ site and not the Te$_{I}$ site. Nonetheless scanning tunneling microscopy (STM) studies have established several  types of defects.\cite{jia2012defects} They are adatoms remnant from the cleavage process, Te$_{I}$-Se$_{II}$ antisites, Te$_{I}$-Bi antisites and  Se$_{I}$-Bi antisites respectively.
A further study using X-ray diffraction (XRD) showed in \BTS{} 8.5\% of Se$_{II}$ atoms are replaced by Te and 4.17 \% of Te$_{I}$ atoms are substituted by Se. The site occupancy factor of Bi is 1.\cite{jia2011low} Therefore, the Se-Te antisite is the major defect in \BTS{}.  A point defect, such as the Te-Se antisite,  can cause a local mode that vibrates at its own frequency with a considerably  broad line-width.\cite{barker1975optical} Such local modes have been observed in  many semiconductors like Ge-Si alloys\cite{feldman1966raman}, CdSe$_{0.985}$S$_{0.015}$,\cite{PhysRev.155.750}, Ba$_{y}$Sr$_{1-y}$F$_{2}$,\cite{PhysRev.164.1169} Sr$_{y}$Ca$_{1-y}$F$_{2}$\cite{PhysRev.164.1169} and so on. A similar temperature dependence of the line-width has been observed  in Iodine intercalated \BSCO.\cite{trodahl1993raman}  There I$_{3}^{+}$ line narrows (19 \wavnum{} to 16 \wavnum{}) from room temperature to 30 K but remains considerably broad  very similar to our observation (14.4 \wavnum{} to 11.7 \wavnum{}). Besides, the intensity of a defect mode can be strong and comparable to other phonon modes as were observed in GaN and Co-doped ZnO films.\cite{siegle1997defect,sudakar2007raman}  Thus, it is very likely the \DM{} is a local mode caused by the point defect Te-Se antisite.

\begin{figure}[!ht]
        \includegraphics[trim=1.5cm 1.5cm 1.5cm 1.5cm,width=\columnwidth]{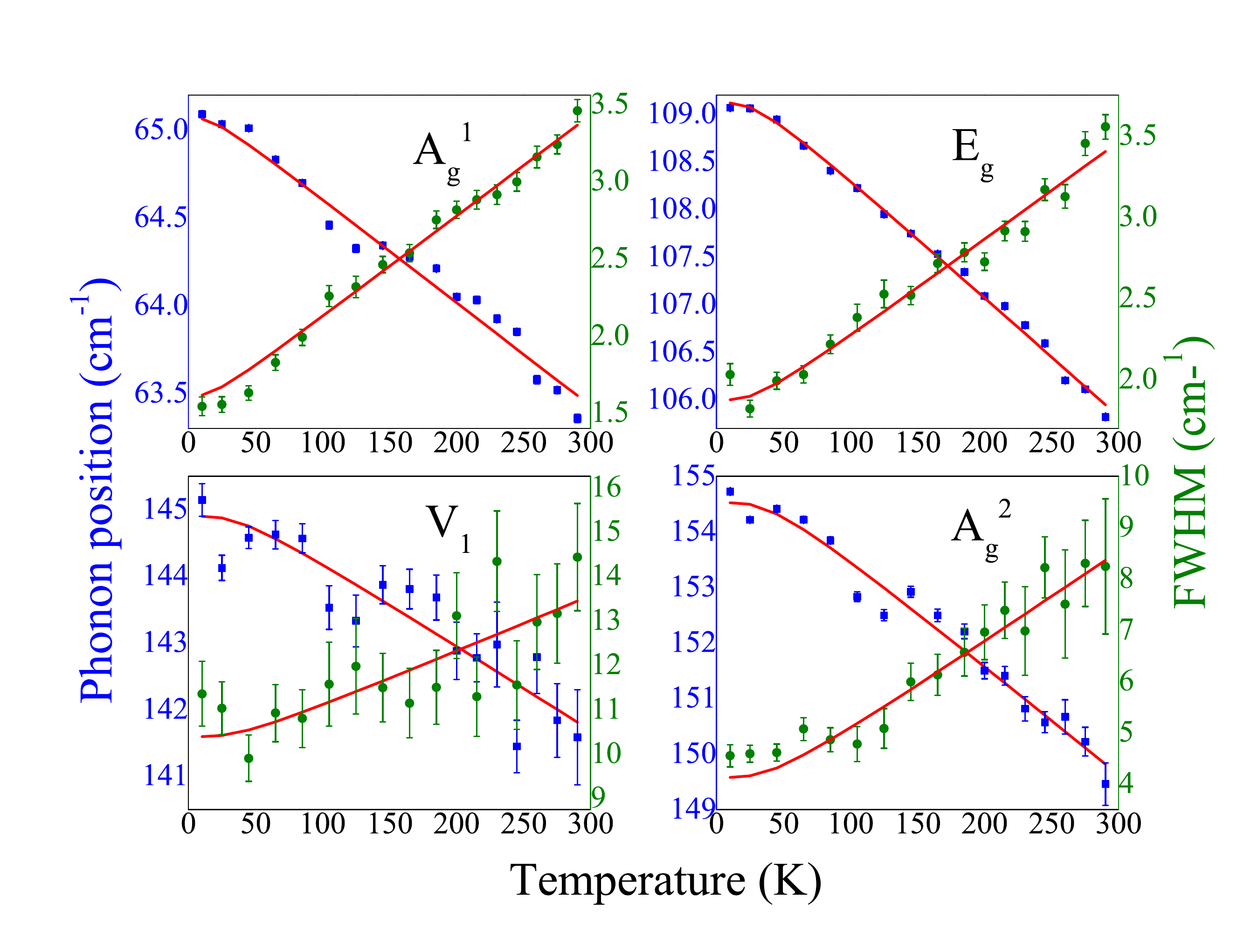}
    \caption{Temperature dependence of the shifts and line-widths of phonons of \BTS.}\label{fig:BTS_Phonon_width}
\end{figure}

To gain more quantitative insights, we focus on
the temperature dependent Raman spectra of \BTS{} over the whole temperature range (shown in FIG. \ref{fig:BTS_raman_colorplot}).
We fit the Raman spectra  with Voigt profile functions,
\begin{equation}\label{voigt_function}
  V(x,\sigma,\Omega,\Gamma)=\int^{+\infty}_{-\infty}G(x',\sigma)L(x-x',\Omega,\Gamma)dx'
\end{equation}
which is the convolution of a Gaussian and a Lorentzian.  The Gaussian is employed to account for the instrumental resolution and the Lorentzian represents a phonon mode. The half width $\sigma$ of the Gaussian was determined by the instrumental resolution, which is 1.8 \wavnum{} in our system.  The extracted temperature dependent phonon shifts $\Omega$ and line-widths $\Gamma$ are shown in FIG. \ref{fig:BTS_Phonon_width}.

\begin{table}
\begin{center}
\begin{tabular}{|rrrrrrrrr|}\hline
 \multicolumn{9}{|c|}{\BTS} \\
 \hline
mode &$\omega_{0}$ &error &C &error& $\Gamma_{0}$ &error &A &error \\ \hline
A$_{g}^{1}$ &   65.3 &     0.1 &    -0.14 &     0.01 &     1.50 &     0.05 &     0.15 &     0.01\\
E$_{g}$ & 109.6 &     0.1 &    -0.50 &     0.01 &     1.64 &     0.08 &     0.24 &     0.02\\
V$_{1}$ & 145.7 &     0.4 &    -0.67 &     0.11 &     9.73 &
0.73 &     0.66 &     0.19\\
A$_{g}^{2}$ &  155.7 &     0.3 &    -1.1 &     0.10 &     3.14 &     0.29 &     1.00 &     0.08\\ \hline
\end{tabular}
\end{center}
\caption{Anharmonic parameters of \BTS{}. The unit is  \wavnum{}.}\label{table:anhamonicity_parameters}
\end{table}

From the plot, we can see that all phonon modes sharpen and harden as the temperature is decreased. This naturally results from the anharmonic phonon-phonon interaction  which leads to a renormalization of  phonon shifts and line-widths. To analyze the temperature dependent behavior,  Klemens's model was used, which is only an approximation to the phonon-phonon interaction, but widely used to interpret anharmonicity.\cite{Raman_Temperature_Bi2Se3,PhysRevB.28.1928} In this model,  an optical phonon is assumed to decay into two phonons with opposite momentum while the ``coalescence'' process where a phonon and a second phonon fuse into a third phonon is neglected, because it requires an existing population of phonons, which are very small at low temperatures.\cite{PhysRevB.50.14179}  Klemens's model reads,\cite{Linear_energyshift_lifetime}
\begin{align}\label{equ:klemens_model_shift}
\Omega(\omega,T)&=\Omega_{0}+C[2n_{B}(\omega_{0}/2)+1]\\
\Gamma(\omega,T)&=\Gamma_{0}+A[2n_{B}(\omega_{0}/2)+1]
\end{align}
where $\Gamma_{0}$ can be interpreted as the contribution from the crystalline disorder, $\Omega_{0}$ is the harmonic phonon shift,  $n_{B}$ is the Bose-Einstein function, and $A$ and $C$  are the coefficients representing three-phonon (cubic) interaction. The resulting fits are also shown in FIG. \ref{fig:BTS_Phonon_width}. Judging for all the figures, we can see that the model works very well. The complete fit parameters including errors are listed in Table \ref{table:anhamonicity_parameters}.
Among all the modes, the temperature independent $\Gamma_{0}$ is dramatically larger for the \DM{}, almost three times that of the A$_{g}^{2}$ mode. However, the $A$ and $C$  follow the trend of increasing as a function of phonon frequency.

To qualitatively explain this, we employed statistical analysis, which was used to handle disordered effects.\cite{PhysRevB.27.858} We can think of \BTS{} crystal as constituted  of many atomic chains.  Raman modes in \BTS{} only involve Bi and Te$_{I}$ sites,\cite{richter1977raman} thus, a perfect chain consists of Bi and Te atoms one after the other. Since our sample has defects, some Te atoms will be randomly replaced by Se atoms. The energies of the local modes in these chains have much larger dependence on the configuration of Se atoms in the chain than that of a normal phonon mode. Consequently, the observed spectra consists of many local modes with slightly different frequencies, resulting in much larger broadening than normal phonon modes.

\begin{figure}
\begin{center}
\includegraphics[width =\columnwidth]{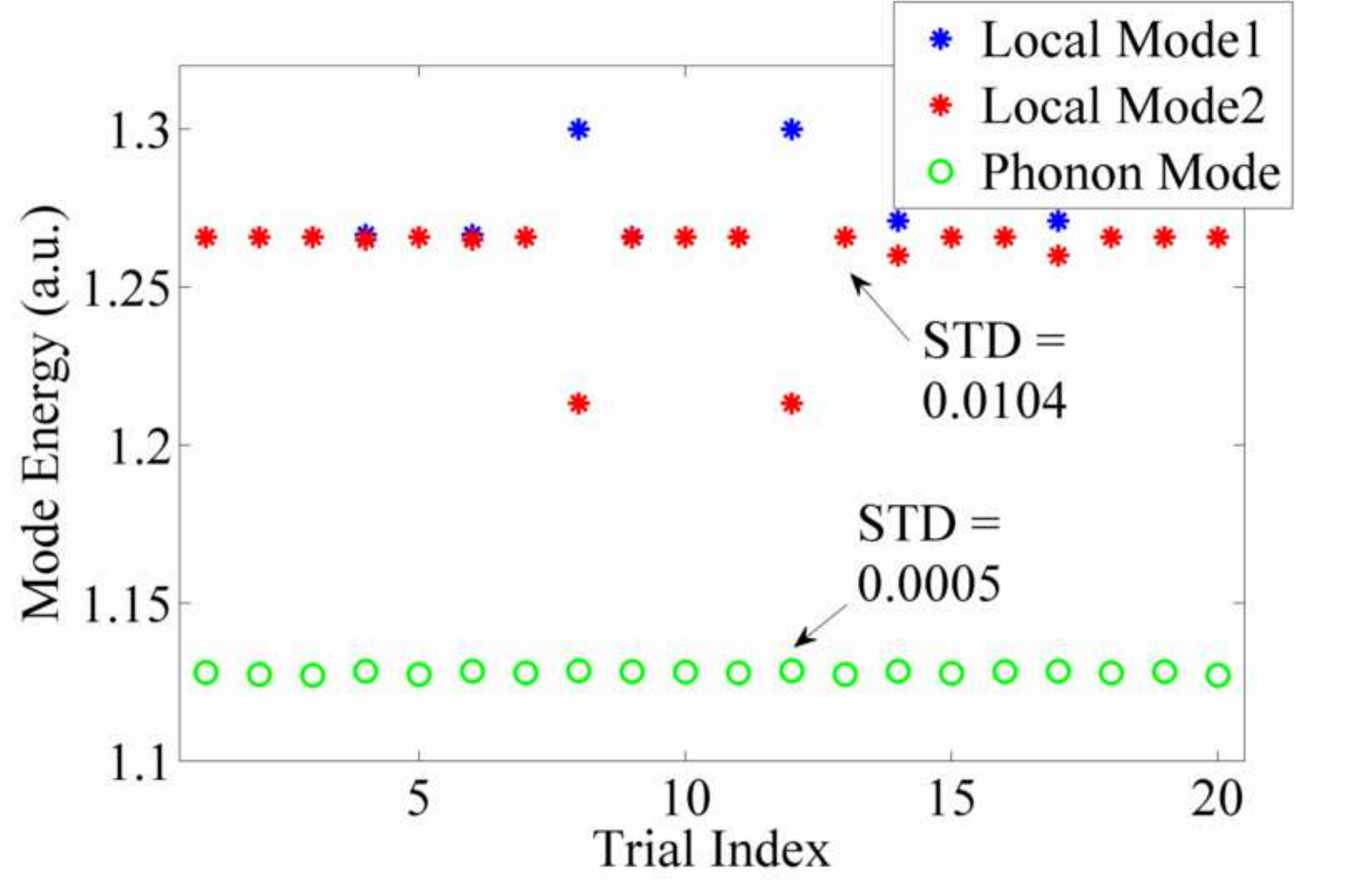}
\end{center}
\caption{Energies of local modes and a phonon mode of the one dimensional circular diatomic chain. The texts denote the corresponding standard deviation (STD).
}
\label{fig:disorder_configuration}
\end{figure}

To test the validity of above statements, a simple model adopted from the literature\cite{barker1975optical} was used to calculate the vibrational modes of a one dimensional circular diatomic chain. The original chain consists of 48 atoms of two types atoms (a,b) with the configuration $...a-b-a-b...$.   The mass of atom $a$ ($b$) was set to 2.58 (4). The force constant was set to be 1. To simulate the local modes, two atoms $a$ were replaced by a ``defect'' atom $c$ with mass 1.57. The masses  were chosen based on the real mass ratio of Te ($a$), Bi ($b$) and Se ($c$) atoms. The number of the ``defect'' atom $c$ was chosen to represent 4.17\% disorder obtained by XRD. Since two atoms are replaced, there will be two local modes in the chain.  Twenty trials were run and for each trial two sites of atom $a$ were randomly chosen to be replaced by atom $c$. The results of two local modes as well as a non-local mode (phonon mode) are shown in FIG. \ref{fig:disorder_configuration}
We can see from the plot, although among different trials, both local modes and the phonon mode are very close in energy, the variation in energy of the local modes is much larger than that of the phonon mode. The other phonon modes were checked as well no qualitative difference was found. Thus our observation that the \DM{} contains a much larger $\Gamma_{0}$ than the other phonon modes is consistent with a mode originating from local defects.

Up to now, we have only discussed the broadening from temperature independent participation of many local modes. For the temperature dependent anharmonic coupling, it has been shown that local modes can couple to lower energy modes as has already been observed in the skutterudites\cite{ogita2008raman}, clathrates\cite{takabatake2014phonon} and other materials.\cite{barker1975optical} As the simulation shows, all the local modes are very close in energy. Therefore, one expects these modes to follow the same temperature dependence it is driven by Bose-Einstein statistics with nearly identical energies.\cite{barker1975optical} Consequently, even though the \DM{} profile in theory consists of many modes, these modes broaden and soften with the same pace as the temperature rises. Thus, the temperature dependent behavior of the \DM{} can still be explained by  Klemens's model.

After clarifying the origin of the anomalous line-width of the \DM{}, we focus on the anharmonic behavior. Comparing the amplitude of `$A$' and `$C$' among the four modes, we find that they monotonically increase with mode energy. This is because the anharmonic coefficients are determined by the phonon joint density of states, $JD(\omega)=\sum\delta(\omega-\omega_{1}-\omega_{2})$. Here $\omega_{1}$ and $\omega_{2}$ are the energies of the two phonons that the phonon $\omega$ decays into. $JD(\omega)$ normally is a monotonically increasing function of $\omega$. Thus the anharmonic coefficients are likely to be a monotonically increasing function of $\omega$ which is consistent with the fit results. 

In summary, we performed  temperature dependent Raman scattering measurement on \BTS{}. Four modes were observed in the entire temperature range. Three of them were assigned as the zone center modes. The extra mode (V$_{1}$) was shown most likely to be an antisite defects induced local mode. The anomalous broadness of the local mode was clarified through an simple model of diatomic chains. The temperature dependence of the four modes were explained by an anharmonic phonon-phonon interaction model. Inelastic Neutron scattering studies on \BTS{} are clearly called for to help elucidate the true nature of this mode and its role in other properties of the material such as thermal transport.
Work at the University of Toronto was supported by NSERC, CFI, and K.S.B. acknowledges support from the National Science Foundation (Grant No. DMR-1410846).


%

\end{document}